\documentclass[conference,twocolumn]{IEEEtran}

\def\paperversion{2} 

\IEEEoverridecommandlockouts

\usepackage{etex}

\usepackage[subtle,indent=normal,mathspacing=normal,wordspacing=normal]{savetrees}

\usepackage{balance}
    
\usepackage[dvipsnames]{xcolor}

\usepackage[cmex10]{amsmath}
\usepackage{amssymb,amsfonts,amssymb,amsthm}
\usepackage{bbm}

\usepackage[T1]{fontenc}
\usepackage{cite}
\usepackage{graphicx}
\usepackage{textcomp}
\usepackage[ruled,vlined]{algorithm2e}
\RestyleAlgo{ruled}
\SetKwComment{Comment}{/* }{ */}
\SetKwInput{KwInput}{Input}
\SetKwInput{KwOutput}{Output}
\SetKw{Continue}{continue}
\SetKw{Break}{break}

\usepackage{url}
\usepackage{enumerate}
\usepackage{float}
\usepackage{booktabs}
\usepackage{multirow}
\usepackage{colortbl}
\usepackage[font=footnotesize]{caption}
\usepackage[font=footnotesize]{subcaption}

\usepackage{cleveref}
\Crefname{figure}{Fig.}{Figs.}

\usepackage{pgfplots}
\usepackage{pgfplotstable}
\usepackage{tikz}
\usetikzlibrary{calc,decorations.pathreplacing,patterns,arrows}

\DeclareCaptionLabelSeparator{periodspace}{.\quad}
\def\BibTeX{{\rm B\kern-.05em{\sc i\kern-.025em b}\kern-.08em
    T\kern-.1667em\lower.7ex\hbox{E}\kern-.125emX}}

\makeatletter
\newcommand{\removelatexerror}{\let\@latex@error\@gobble}
\makeatother

\DeclareMathOperator*{\argmax}{arg\,max}
\newcommand{\tmax}{{t_{\text{max}}}}
\newcommand{\tavg}{{t_{\text{avg}}}}

\begin{document}

\title{Time Efficiency of BATS Coding on Wireless Relay\\Network With Overhearing}
\author{%
  \IEEEauthorblockN{Hoover~H.~F.~Yin}
  \thanks{H.~H.~F.~Yin is with the Department of Information Engineering, The Chinese University of Hong Kong.}
}

\maketitle

\begin{abstract}
Wireless relay network is a solution to extend the reach of a wireless connection by installing a relay node between the source node and the sink node.
Due to the broadcast nature of wireless transmission, the sink node has a chance to receive part of the data sent by the source node.
In this paper, we apply a network coding scheme called BATS codes on a wireless relay network where the relay node has a stable power supply, so that we can aim for the best decoding time instead of minimizing the number of transmissions for saving energy.
We optimize the time efficiency that maximize the average decoding rate per unit time by some heuristics, and bring out a message that 
it is not optimal to set an average number of recoded packets per batch at the relay node equals the number of packets per batch sent by the source node.
\end{abstract}

\section{Introduction}

In a common wireless network, the traffic generated by an end-user first travels through a wireless link to reach an access point (AP), and then follows by wired links to reach the Internet.
Due to signal fading or blockage of line-of-slight, the packet loss rate between the user and the AP can be large.
A simple solution is to install a wireless relay between the user and the AP to form a wireless relay network.

Due to the broadcast nature of wireless transmission, the sink node (AP) has a chance to receive the packets sent by the source node (user) to the relay node.
In other words, there is an implicit overhearing channel in a wireless relay network.
The relay node does not need to send the duplicated information that the sink node has already been received, thus we can have a better utilization of the channel from the relay node to the sink node.
Traditionally, the relay node acts as a repeater that forwards the signal it receives.
With network coding \cite{linear,flow}, the relay node acts as a \emph{recoder} that performs packet combining operations. 
In general, network coding has throughput gain over forwarding.

\emph{Batched sparse (BATS) code} \cite{yang14bats,bats_book} is a variant of \emph{random linear network coding (RLNC)} \cite{random2} that generates small subsets of coded packets called \emph{batches}, and then restricts the application of RLNC to the packets belonging to the same batch.
Among other batch-wise approaches \cite{silva09,heidarzadeh10,li11,tang12,mahdaviani12,tang18}, BATS code has the best achievable rate.

To enhance the throughput, we need to decide the number of recoded packets for each batch according to the receiving status of that batch \cite{yang14a}.
If we fix the average number of recoded packets per batch, $\tavg$, the problem can be solved by 
an optimization known as \emph{adaptive recoding} \cite{adaptive,scheduling,ge_adaptive,uni,bar}.
Yet, $\tavg$ is another parameter to be optimized.
In \cite{zhang2017efficient,rf}, $\tavg$ is optimized by minimizing the total number of packets to be sent by both the source and the relay nodes.
The corresponding measure is called the \emph{packet efficiency} or the \emph{energy efficiency} as each transmission consumes energy, which is an important factor for IoT devices that have limited power supply, e.g., run on batteries.
The time required to decode the data is not the top priority in this setting.

In this paper, we consider the other end of the spectrum when the relay node has a stable (unlimited) power supply, and
aim for optimizing the decoding time.
More specifically, we consider the \emph{time efficiency} that measures the average number of decoded input packets per unit time.
At first glance, one may think that the average number of recoded packets per batch, $\tavg$, should be the same as the number of packets per batch sent by the source node.
However, the effect of idling time at the relay node is not considered:
When the relay node has no more batch to recode, an idling period is introduced until the source node finishes sending the next batch.
This way, the time for the relay to finish sending all batches is deferred.

We want to bring out a message in this paper that, the ``first glance'' is incorrect instead.
From our numerical evaluation, $\tavg$ is smaller than the number of packets per batch sent by the source node.
We also discuss some heuristics to optimize the zigzag-shaped time efficiency model, which is hard to solve in general.

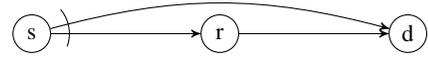
\begin{figure}
    \centering
    \small
    \begin{tikzpicture}
        \node[draw,circle,inner sep=0pt,minimum size=.5cm] (n0) at (0,0) {s};
        \node[draw,circle,inner sep=0pt,minimum size=.5cm] (n1) at (2.5,0) {r};
        \node[draw,circle,inner sep=0pt,minimum size=.5cm] (n2) at (5,0) {d};
        \draw[->,>=stealth'] (n0) -- (n1);
        \draw[->,>=stealth'] (n1) -- (n2);
        \draw[->,>=stealth'] (n0) to [out=15,in=165] (n2);
        \draw[domain=-20:40] plot ({.5*cos(\x)}, {.5*sin(\x)});
    \end{tikzpicture}
    \caption{A wireless relay network.
        The source node (s) broadcasts the packets to the relay node (r) and the destination node (d).
        The link between (s) and (d) is considered as an overhearing channel.
    }
    \label{fig:net}
\end{figure}

\begin{table}
\vskip 1em
    \let\oldtabcolsep\tabcolsep
    \setlength{\tabcolsep}{4pt}
    \centering
    \caption{Table of Notations}
    \label{tab:notations}
    \vskip -0.7em
    \scriptsize
    \begin{tabular}{cl|cl}
        \toprule
        $M$ & batch size & $\omega$ & time for 1 tx. at source\\
        $F$ & \# input pkts. & $B$ & \# of batches sent by source\\
        $\{h_r\}$ & innov. rank dist. & $E$ & exp. rank at sink (i.e., throughput)\\
        $E(r,t)$ & exp. rank func. & $R$ & exp. rank at sink before relay starts tx.\\
        $D$ & exp. total idling time & $\tavg$ & avg. \# of recoded pkt. per batch\\
        $\Delta_\tavg$ & $\max_{r = 0}^M \Delta_{r, \lfloor t_r \rfloor}$ & $\Delta_{r,t}$ & $E(r,t+1)-E(r,t)$\\
    \end{tabular}
    \setlength{\tabcolsep}{\oldtabcolsep}
\end{table}

\section{BATS Codes}

We apply BATS codes on the wireless relay network shown in \Cref{fig:net} to send a piece of data from the source node to the sink node.
We assume constant transmission rates at the source and the relay nodes with the following normalization:
Each transmission at the relay node takes $1$ time unit, and that at the source node takes $\omega$ time units.
Some notations that are used throughout the paper are summarized in \Cref{tab:notations} for quick reference.

\subsection{Encoding, Recoding and Decoding}

The data to be sent is partitioned into a set of \emph{input packets}.
Each input packet is a vector of the same length over a fixed finite field $\mathbb{F}$, where this length can be optimized to reduce padding overheads \cite{pktsize}.
To generate a batch, the encoder first samples a ``batch degree'' from a predefined degree distribution.
To achieve a close-to-optimal rate, this distribution must be optimized \cite{yang14bats,tree,deg_dro,quasi}.
According to the 
degree, a subset of all input packets is chosen uniform randomly.
After that, random linear combinations on the chosen packets are performed to generate the coded packets of the batch.

The number of packets in a freshly generated batch, denoted by $M$, is called the \emph{batch size}.
These $M$ packets are defined to be linearly independent of each other, which can be achieved by proper initialization of coefficient vectors \cite{yang22pro,yin20pro}.
Two packets in the same batch are said to be linearly independent of each other if and only if their coefficient vectors are linearly independent of each other.
The number of linearly independent packets in a batch is called the \emph{rank} of the batch.
The packets of the batch are then sent to the relay node, while the sink node may overhear some of these packets.

At the relay node, we can start \emph{recoding} a batch after the source node finishes sending the packets of this batch.
The number of recoded packets to be generated can be obtained via \emph{adaptive recoding} \cite{adaptive,scheduling,ge_adaptive,uni,bar} that 
maximizes the expected rank of the batches arriving at the sink node.
This expected rank is the theoretical upper bound on the achievable rate \cite{yang11x2}.

At the sink node, belief propagation algorithm and inactivation decoding \cite{shokRaptor,Raptormono} can be applied together to decode the batches.
In an optimized BATS code, each received rank at the sink node almost corresponds to an input packet.
Therefore, we have the approximation
    $F/B \approx E$,
where $F$ is the number of input packets of the data at the source node, $B$ is the number of batches sent by the source node, and $E$ is the expected rank of the batches arriving at the sink node \cite{zhou17b}.
Thus, the number of batches to ensure decodability can be approximated by $\lceil F/E \rceil$.

\subsection{Adaptive Recoding}

For each batch, let $V$ and $V'$ be two vector subspaces of $\mathbb{F}^M$ that are spanned by the coefficient vectors of the packets in this batch received by the relay and the sink nodes respectively.
The \emph{innovative rank} of the batch is defined as $\dim(V+V') - \dim(V')$.
In other words, the innovative rank measures the remaining ``ranks'' the relay node can provide the sink node.
Denote by $\mathbf{h} = (h_0, \ldots, h_M)^T$ the innovative rank distribution, where its formula can be found in \cite{innov}.

Denote by $E(r,t_r)$ the increment of the rank of a batch at the sink node conditioning on that at the relay node, this batch has innovative rank $r$, and $t_r$ recoded packets are being sent.
Its formulation depends on the packet loss model.
For example, for independent loss rate $p$ with a sufficiently large field, we have
\begin{equation*}
    E(r,t_r) = \sum_{i = 0}^{t_r} \binom{t_r}{i} (1-p)^i p^{t_r-i} \min\{i, r\}.
\end{equation*}
This $E(r,t_r)$ is monotonically increasing and concave with respect to $t_r$ \cite{uni}.
For simplicity, we denote $\Delta_{r,t} = E(r,t+1)-E(r,t)$ for non-negative integer $t$ in the algorithms in this paper,

When $t_r \ge 0$ is not an integer, we define $E(r,t)$ by linear interpolation. 
That is, we first send $\lfloor t_r \rfloor$ packets.
The fractional part of $t_r$ is the probability to send one more packet.
The expected rank of the batches arriving at the sink node can be expressed as
\begin{equation*}
    E = R + \sum_{r = 0}^M h_r E(r, t_r),
\end{equation*}
where $R$ is the expected rank at the sink node without considering the packets sent by the relay node \cite{rf,innov}.
Let $\tavg$ be the average number of recoded packets per batch.
Define $\mathbf{t} = (t_0, t_1, \ldots, t_M)^T$, where $t_i$ is the number of recoded packets for a batch of innovative rank $i$.
The adaptive recoding optimization is the concave problem
\begin{equation}
\label{eq:tap}
    \max_{\mathbf{t} \ge \mathbf{0}}\, R + \sum_{r = 0}^M h_r E(r, t_r) \quad \mathrm{s.t.} \quad \mathbf{h}^T \mathbf{t} = \tavg,
\end{equation}
which can be solved by the greedy algorithm in \cite{uni}.

\section{Time Efficiency}

Suppose the relay node sends an average of $\tavg$ packets per batch, and both source node and relay node can send packets at the same time.
Note that the time required for the relay node to send all $B$ batches is not as simple as $B\tavg$ time units.

First, the relay node needs to receive some packets from the source node before it can start generating recoded packets.
If the relay node starts recoding a batch after the source node finishes sending the packets of this batch, then there is a delay of $\omega M$ time unit at the beginning.
This initial delay only occurs once, as the starting time of recoding for all batches at the relay node are shifted by this delay.

Second, if the relay node sends too few packets for a batch, the source node may not finish the transmission of the next batch yet.
This way, some idling timeslots are introduced to the relay node before it can start recoding the next batch.
Yet, these idling timeslots may be reduced if the relay node has not finished sending the recoded packets of the previous batch.

Denote by $D$ the expected total idling time at the relay node.
The time required for the relay node to send all $B$ batches is $B\tavg + D$ time units.
That is, the time required for the sink node to receive enough batches for decoding is $\max\{B \omega M, B\tavg + D\}$ time units.
This can be interpreted as the decoding time as $B$ is already minimized by the optimal degree distribution.
The \emph{time efficiency} is defined as
\begin{IEEEeqnarray*}{Cl}
    & \frac{F}{\max\{B \omega M, B\tavg + D\}} \approx \frac{R + \sum_{r = 0}^M h_r E(r, t_r)}{\max\{\omega M, \tavg + D/B\}}\\
    \approx & \frac{R + \sum_{r = 0}^M h_r E(r, t_r)}{\max\{\omega M, \tavg + \frac{D}{F}(R + \sum_{r = 0}^M h_r E(r, t_r))\}}.
\end{IEEEeqnarray*}
The optimization problem is then
\begin{equation} \tag{E0} 
\label{eq:eff}
    \begin{IEEEeqnarraybox}[][c]{rCl}
        \max_{\mathbf{t} \ge \mathbf{0}, \tavg \ge 0} & \,\,\, & \frac{R + \sum_{r = 0}^M h_r E(r, t_r)}{\max\{\omega M, \tavg + \frac{D}{F}(R + \sum_{r = 0}^M h_r E(r, t_r))\}}\\
        \mathrm{s.t.} && \mathbf{h}^T \mathbf{t} = \tavg.
    \end{IEEEeqnarraybox}
\end{equation}
Maximizing the time efficiency is equivalent to minimizing the decoding time.
However, we can exploit the structure of time efficiency to partially reduce it to a simpler problem (see \Cref{sec:alg}).

\subsection{Expected Total Idling Time}

Let $\bar{t}_i$ be the probability of sending $i$ recoded packets for a batch at the relay node.
That is,
\begin{equation*}
    \bar{t}_i = \sum_{r = 0}^M h_r  \left( (1-(t_r - \lfloor t_r \rfloor))\mathbbm{1}_{\lfloor t_r \rfloor = i} + (t_r - \lfloor t_r \rfloor)\mathbbm{1}_{\lfloor t_r \rfloor + 1 = i} \right)
\end{equation*}
where $\sum_{i = 0}^\infty \bar{t}_i = 1$, and $\mathbbm{1}$ is the indicator function.
Next, let $Q_b$ be a random variable for the time the source node finishes sending the $(b+1)$-th batch minus the time the relay node finishes sending the $b$-th batch.
If $Q_b > 0$, then the relay node has nothing to send until the source node finishes sending the $(b+1)$-th batch, i.e., there is an idling period of length $\omega M - Q_b$ time units.
If $Q_b \le 0$, then there is no idling time as the relay node can start sending the recoded packets of the $(b+1)$-th batch.
We have the Markov process
\begin{equation*}
    \Pr(Q_{b+1} = \min\{q, 0\} + \omega M - i \mid Q_b = q) = \bar{t}_i.
\end{equation*}

In the formulation of time efficiency, we need the expected total idling time at the relay node.
This can be expressed as
\begin{equation} \label{eq:D_rec}
    \textstyle D = \omega M + \mathbb{E}\left[\sum_{b = 1}^{B-1} (M - Q_b) \mathbbm{1}_{Q_b > 0}\right],
\end{equation}
where the extra $\omega M$ term is the initial delay.
We sum up to $B-1$ batches as there is no idling time needed after sending the last batch.
An efficient and practical way to approximate $D$ is to use the Monte Carlo method.
\ifnum\paperversion=1
However, if $\omega$ is rational, it is possible to write an analytical form of $D$.
See \cite{rf2_arxiv} for an example.
\fi
\ifnum\paperversion=2
However, if $\omega$ is rational, then we can form a Markov chain for the process $\{Q_b\}$ with countable number of states.
This way, it is possible to write an analytical form of $D$.
\fi

\ifnum\paperversion=2
We take $\omega = 1$ as an example to illustrate the analytical form of the expected total idling time when the number of states is countable.
The transition matrix of the Markov chain is an infinite matrix
\begin{equation*}
    \mathbf{P} = \begin{pmatrix}
        \bar{t}_0 & \bar{t}_1 & \bar{t}_2 & \cdots\\
        \bar{t}_0 & \bar{t}_1 & \bar{t}_2 & \cdots\\
        \vdots & \vdots & \cdots & \ddots\\
        \bar{t}_0 & \bar{t}_1 & \bar{t}_2 & \cdots\\
        0 & \bar{t}_0 & \bar{t}_1 & \cdots\\
        0 & 0 & \bar{t}_0 & \cdots\\
        \vdots & \vdots & \cdots & \ddots
    \end{pmatrix}
\end{equation*}
where the first $M+1$ rows are identical.
The $1$-st row corresponds to the state $M$, i.e., $M$ idling timeslots.
The $2$-nd row corresponds to the state $M-1$, and so on and so forth.
The expected idling time after sending the $b$-th batch is
\begin{equation*}
    (1, 0, 0, \ldots) \mathbf{P}^b (M, M - 1, M - 2, \ldots, 1, 0, 0, 0, \ldots)^T.
\end{equation*}
Although the initial state is $0$ (to avoid counting extra idling timeslots), the corresponding row in $\mathbf{P}$ is the same as the $1$-st row, thus we can set the initial probability vector as $(1, 0, 0, \ldots)$.
The expected total idling time is therefore
\begin{equation*}
    M + (1, 0, 0, \ldots) \sum_{b = 1}^{B-1} \mathbf{P}^b (M, M - 1, M - 2, \ldots, 1, 0, 0, 0, \ldots)^T.
\end{equation*}
For numerical calculation, we can truncate $\mathbf{P}$ into a matrix of finite size, as it is unusual to send too many recoded packets for a batch.
This assumption is common in literature such as \cite{scheduling,ge_adaptive,wang2021smallsample}.

It is worth to remark that when $B$ is large, we cannot approximate $\sum_{b = 1}^{B-1} \mathbf{P}^b$ by setting $B \to \infty$.
If $\sum_{b = 1}^\infty \mathbf{P}^b$ converges, then it equals $(\mathbf{I} - \mathbf{P})^{-1}$.
The necessary and sufficient condition is that spectral radius of $\mathbf{P} < 1$.
However, $\mathbf{P}$ is a stochastic matrix so its largest eigenvalue is $1$.
This means that the geometric sum does not converge.
\fi

\begin{figure}
\centering
\removelatexerror
\begin{algorithm}[H]
\small
  \SetKwInOut{Input}{Input}
  \SetKwInOut{Output}{Output}
  \ResetInOut{Output}
  \Input{\# of input packets $F$, batch size $M$, \# of batches $B$,\\innov. rank dist. $\mathbf{h}$, initial expected rank $R$}
  \Output{time efficiency and $\tavg$ at the end-points of the segment} 
  $\tavg \gets 0$ ; $\mathbf{t} \gets \mathbf{0}$ ; $E \gets R$ \;
  \While(\tcp*[f]{embedded greedy algo. in \cite{uni}}){True}{ 
    $\Delta_\tavg \gets \max_{r = 0}^M \Delta_{r,t_r}$ \;
    $r' \gets$ an element in $\argmax_{r = 0}^M \Delta_{r,t_r}$ \;
    $s \gets (F/B - E)/\Delta_\tavg$ \;
    \lIf{$s \le h_{r'} (1 - (t_{r'} - \lfloor t_{r'} \rfloor))$}{
        \Break
    }
    $s \gets h_{r'} (1 - (t_{r'} - \lfloor t_{r'} \rfloor))$ \;
    $t_{r'} \gets t_{r'} + 1$ ; $\tavg \gets \tavg + s$ \;
    $E \gets E + s\Delta_\tavg$ \;
  }
  $t_{r'} \gets t_{r'} + s/h_{r'}$ ; $\tavg \gets \tavg + s$ \;
  $E \gets E + s\Delta_\tavg$ \;
  $D \gets$ compute the expected idling time (Eq.~\eqref{eq:D_rec}) \;
  $e_\ell \gets \left(\tavg, \min\left\{\frac{E}{\tavg+D/B}, \frac{E}{\omega M}\right\}\right)$ \;
  \While(\tcp*[f]{embedded greedy algo. in \cite{uni}}){True}{
    $\Delta_\tavg \gets \max_{r = 0}^M \Delta_{r,t_r}$ \;
    $r' \gets$ an element in $\argmax_{r = 0}^M \Delta_{r,t_r}$ \;
    $s \gets (F/(B-1) - E)/\Delta_\tavg$ \;
    \lIf{$s \le h_{r'} (1 - (t_{r'} - \lfloor t_{r'} \rfloor))$}{
        \Break
    }
    $s \gets h_{r'} (1 - (t_{r'} - \lfloor t_{r'} \rfloor))$ \;
    $t_{r'} \gets t_{r'} + 1$ ; $\tavg \gets \tavg + s$ \;
    $E \gets E + s\Delta_\tavg$ \;
  }
  $t_{r'} \gets t_{r'} + (s-\epsilon)/h_{r'}$ ; $\tavg \gets \tavg + (s-\epsilon)$ \;
  $E \gets E + (s-\epsilon)\Delta_\tavg$ \;
  $D \gets$ compute the expected idling time (Eq.~\eqref{eq:D_rec}) \;
  $e_r \gets \left(\tavg, \min\left\{\frac{E}{\tavg+D/B}, \frac{E}{\omega M}\right\}\right)$ \;
  \KwRet $(e_\ell, e_r)$ \tcp*{the 2 end-point coord.}
  \caption{End-Points of the Time Efficiency Segment} 
  \label{alg:solve}
\end{algorithm}
\vskip -.5em
\end{figure}

\subsection{Algorithm}
\label{sec:alg}

We first fix $\tavg$ and then split \eqref{eq:eff} into two subproblems:
\begin{equation} \tag{E1} 
\label{eq:eff0}
    \max_{\mathbf{t} \ge \mathbf{0}} \frac{R + \sum_{r = 0}^M h_r E(r, t_r)}{\omega M} \quad \mathrm{s.t.} \quad \mathbf{h}^T \mathbf{t} = \tavg,
\end{equation}
\begin{equation} \tag{E2} 
\label{eq:eff1}
    \max_{\mathbf{t} \ge \mathbf{0}} \frac{R + \sum_{r = 0}^M h_r E(r, t_r)}{\tavg + \frac{D}{F}(R + \sum_{r = 0}^M h_r E(r, t_r))} \,\,\, \mathrm{s.t.} \,\,\, \mathbf{h}^T \mathbf{t} = \tavg.
\end{equation}
The minimum of the their solutions is the solution to \eqref{eq:eff} for the fixed $\tavg$.
The first problem is the concave problem \eqref{eq:tap}, where the objective increases when $\tavg$ increases.
So, the remaining problems are to solve \eqref{eq:eff1} and search for the optimal $\tavg$.
Although the numerator of the objective of \eqref{eq:eff1} is concave, we cannot conclude whether its denominator is convex or not.
The reason is that the value of $D$ depends on $B$ that depends on $R + \sum_{r = 0}^M h_r E(r, t_r)$.
If $B$ is changed when we increase $\tavg$, the new $D$, denoted as $D'$, is possible to be $D' \ge D$.

Note that when we increase $\tavg$ as long as $B$ is unchanged, we are utilizing more idling timeslots, thus $D'$ is monotonically decreasing.
From our numerical evaluation, we observe that the objective of \eqref{eq:eff1} is almost linear when we increase $\tavg$ while keep $B$ unchanged.
As a heuristic, for each $B$, we can test the two ends of the ``line segment'', i.e., the smallest and largest $\tavg$ such that $B$ remains the same, to find a suboptimal time efficiency.
One of the segments on the two sides of this suboptimal point contains the optimal time efficiency, so we can later perform a search on this restricted interval of $\tavg$.

\Cref{alg:solve} is a realization of the above discussion, with the greedy algorithm in \cite{uni} embedded (to use the internal state of the algorithm) for searching the values of $\tavg$ that are the end-points of the segment for a given $B$.
The first half of the algorithm finds the smallest $\tavg$ that corresponds to sending $B$ batches, i.e., the left end-point, denoted by $e_\ell$.
The second half finds the right end-point, denoted by $e_r$.
Yet, we cannot reach the right end-point precisely because this will change the value of $B$.
Therefore, we consider a sufficiently small $\epsilon > 0$ from the right end-point of the segment.

After identifying the range of $\tavg$ by \Cref{alg:solve}, we consider a discretized set of this interval with a sufficiently small step size for finding the optimal time efficiency as shown in \Cref{alg:solve2}.

\begin{figure}
\centering
\removelatexerror
\begin{algorithm}[H]
\small
  \SetKwInOut{Input}{Input}
  \SetKwInOut{Output}{Output}
  \ResetInOut{Output}
  \Input{\# of input packets $F$, batch size $M$, innov. rank dist. $\mathbf{h}$,\\initial expected rank $R$, an interval $[a, b]$ for searching $\tavg$}
  \Output{optimal time efficiency for $\tavg \in [a, b]$}
  $f^\ast \gets 0$ ; $\tavg \gets a$ ; $\mathbf{t} \gets \mathbf{0}$ ; $E \gets R$ ; $s \gets \tavg$ \;
  \While{$\tavg \le b$}{
  \While(\tcp*[f]{embedded greedy algo.}){$s > 0$}{
    $\Delta_\tavg \gets \max_{r = 0}^M \Delta_{r,t_r}$ \;
    $r' \gets$ an element in $\argmax_{r = 0}^M \Delta_{r,t_r}$ \;
    \uIf{$h_{r'} \ge s$}{
        $t_{r'} \gets t_{r'} + s/h_{r'}$ ;
        $E \gets E + s\Delta_\tavg$ ;
        $s \gets 0$ \;
    }\Else{
        $t_{r'} \gets t_{r'} + 1$ ;
        $E \gets E + \Delta_\tavg$ ;
        $s \gets s - h_{r'}$ \;
    }
  }
  $B \gets \lceil F/E \rceil$ \;
  $D \gets$ compute the expected idling time (Eq.~\eqref{eq:D_rec}) \;
  $f \gets \min\left\{\frac{E}{\tavg+D/B}, \frac{E}{\omega M}\right\}$ \;
  \lIf{$f > f^\ast$}{
    $f^\ast \gets f$ 
  }
  $s \gets \epsilon$ ; $\tavg \gets \tavg + \epsilon$ \;
  }
  \KwRet $f^\ast$ \;
  \caption{Search for Optimal Time Efficiency} 
  \label{alg:solve2}
\end{algorithm}
\vskip -.5em
\end{figure}

\subsection{Upper Bound}

When $F$ is larger, $B$ is also larger, and $D/B$ tends to be smaller.
For large $F$, we may approximate $D/B \to 0$. 
Then, \eqref{eq:eff} becomes the following concave-convex fractional programming problem that represents an upper bound on the time efficiency:
\begin{equation*}
    \max_{\mathbf{t} \ge \mathbf{0}, \tavg \ge 0} \frac{R + \sum_{r = 0}^M h_r E(r, t_r)}{\max\{\omega M, \tavg\}} \quad \mathrm{s.t.} \quad \mathbf{h}^T \mathbf{t} = \tavg.
\end{equation*}
\ifnum\paperversion=1
This can be rewritten as a linear program. 
See \cite{rf2_arxiv} for details.
\fi
\ifnum\paperversion=2
By Schaible transform, we get the concave problem
\begin{equation*}
    \begin{IEEEeqnarraybox}[][c]{rCl}
        \max_{\mathbf{t}' \ge \mathbf{0}, \tavg' \ge 0, s \ge 0} & \quad & s \left( R + \sum_{r = 0}^M h_r E(r, t'_r/s) \right)\\
        \mathrm{s.t.} && s \max\{\omega M, \tavg'/s\} \le 1, \quad \mathbf{h}^T \mathbf{t}' = \tavg',
    \end{IEEEeqnarraybox}
\end{equation*}
where $\tavg = \tavg'/s$ and $\mathbf{t} = \mathbf{t}'/s$.

Although the perspective of the concave $E(r, t_r)$, i.e., $s E(r, t_r/s)$, is concave, optimization solvers may not be able to detect it.
If we fix an integral upper bound $\tmax$ on $t_r$ for all $r$, then we can linearize $E(r, t_r)$ by
\begin{equation*}
    E(r, t_r) = \min_{i \in \{0, 1, \ldots, \tmax\}} (E(r, i) + (t_r - i) \Delta_{r,i})
\end{equation*}
due to its linear interpolation nature \cite{wang2021smallsample}.
The problem becomes a linear programming problem:
\begin{equation*}
    \begin{IEEEeqnarraybox}[][c]{rCl}
        \max_{\mathbf{t}' \ge \mathbf{0}, \tavg' \ge 0, s \ge 0, \mathbf{u}'} & \quad & sR + \mathbf{h}^T \mathbf{u}\\
        \mathrm{s.t.} && u'_r \le sE(r, i) + (t'_r - si) \Delta_{r,i},\\
        && \qquad\qquad \forall i \in \{0, 1, \ldots, \tmax\},\\
        && s \omega M \le 1, \quad \tavg' \le 1, \quad \mathbf{h}^T \mathbf{t}' = \tavg'.
    \end{IEEEeqnarraybox}
\end{equation*}
\fi


\section{Numerical Evaluation}

\begin{figure}
    \centering
    \begin{subfigure}{.225\textwidth}
    \centering
    \begin{tikzpicture}
        \begin{axis}[
            small,
            scale=0.7,
            xtick={1,2,...,10},
            ytick={0.0,0.1,...,0.9},
            xmin=0.01, xmax=10,
            ymin=0.0, ymax=0.9,
            xlabel=$\tavg$,
            ylabel=time efficiency,
            x label style={at={(axis description cs:.5,.13)},anchor=north},
            y label style={at={(axis description cs:.25,.5)},anchor=south},
            label style={font=\scriptsize},
            tick label style={font=\tiny},
            legend style={legend columns=4,font=\scriptsize,at={(-0.1,1.05)},anchor=south west}
        ]
            \node[] at (axis cs: 8,0.1) {\scriptsize $M = 8$};
            \addplot[black,dashed] table [x=tavg, y=eff2, col sep=comma] {data/all_100_8_8_20_20_80.csv};
            \addlegendentry{\eqref{eq:eff0}}
            \addplot[red] table [x=tavg, y=eff1, col sep=comma] {data/all_100_8_8_20_20_80.csv};
            \addlegendentry{\eqref{eq:eff1}, $F = 100$}
            \addplot[blue] table [x=tavg, y=eff1, col sep=comma] {data/all_256_8_8_20_20_80.csv};
            \addlegendentry{\eqref{eq:eff1}, $F = 256$}
            \addplot[cyan] table [x=tavg, y=eff1, col sep=comma] {data/all_512_8_8_20_20_80.csv};
            \addlegendentry{\eqref{eq:eff1}, $F = 512$}
        \end{axis}
    \end{tikzpicture}
    \vskip -.5em
    \end{subfigure}
    \hspace{5pt}
    \begin{subfigure}{.225\textwidth}
    \centering
    \begin{tikzpicture}
        \begin{axis}[
            small,
            scale=0.7,
            xtick={2,4,...,18},
            ytick={0.0,0.1,...,0.9},
            xmin=0.01, xmax=18,
            ymin=0.0, ymax=0.9,
            xlabel=$\tavg$,
            x label style={at={(axis description cs:.5,.13)},anchor=north},
            y label style={at={(axis description cs:.25,.5)},anchor=south},
            label style={font=\scriptsize},
            tick label style={font=\tiny},
        ]
            \node[] at (axis cs: 14,0.1) {\scriptsize $M = 16$};
            \addplot[black,dashed] table [x=tavg, y=eff2, col sep=comma] {data/all_100_16_16_20_20_80.csv};
            \addplot[red] table [x=tavg, y=eff1, col sep=comma] {data/all_100_16_16_20_20_80.csv};
            \addplot[blue] table [x=tavg, y=eff1, col sep=comma] {data/all_256_16_16_20_20_80.csv};
            \addplot[cyan] table [x=tavg, y=eff1, col sep=comma] {data/all_512_16_16_20_20_80.csv};
        \end{axis}
    \end{tikzpicture}
    \vskip -.5em
    \end{subfigure}
    \vskip -.7em
    \caption{The time efficiency achieved by the objectives of \eqref{eq:eff0} and \eqref{eq:eff1}.} 
    \label{fig:all_eff}
    \vskip -.5em
\end{figure}
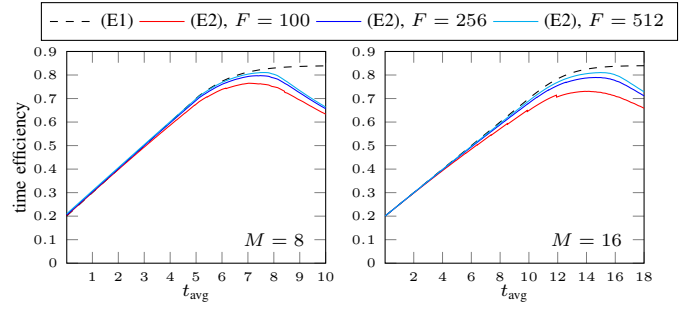

We evaluate the objectives of \eqref{eq:eff0} and \eqref{eq:eff1}.
Consider $20\%$ single-hop and $80\%$ double-hop packet loss rates in a wireless relay network.
\Cref{fig:all_eff} illustrates the objectives when $F \in \{100, 256, 512\}$. 
The dashed curve is the objective of \eqref{eq:eff0}, which is independent of $F$.
The objective in \eqref{eq:eff} is the minimum of a colored curve and the dashed curve.
In general, the objective of \eqref{eq:eff1} becomes smaller than that of \eqref{eq:eff0} when $\tavg$ increases, thus it is necessary to evaluate the complicated formulation in \eqref{eq:eff1}.

\begin{figure}
    \centering
    \begin{subfigure}{.225\textwidth}
    \centering
    \begin{tikzpicture}
        \begin{axis}[
            small,
            scale=0.7,
            xtick={1,1.5,2,...,10},
            ytick={0.0,0.01,...,0.9},
            xmin=6, xmax=9,
            ymin=0.7, ymax=0.77,
            xlabel=$\tavg$,
            ylabel=time efficiency,
            x label style={at={(axis description cs:.5,.13)},anchor=north},
            y label style={at={(axis description cs:.22,.5)},anchor=south},
            label style={font=\scriptsize},
            tick label style={font=\tiny},
            legend style={legend columns=2,font=\scriptsize,at={(0,1.05)},anchor=south west}
        ]
            \node[] at (axis cs: 6.5,0.713) {\scriptsize $F = 100$};
            \node[] at (axis cs: 6.5,0.707) {\scriptsize $M = 8$};
            \addplot[red] table [x=tavg, y=eff1, col sep=comma] {data/all_100_8_8_20_20_80.csv};
        \end{axis}
    \end{tikzpicture}
    \vskip -.5em
    \end{subfigure}
    \hspace{5pt}
    \begin{subfigure}{.225\textwidth}
    \centering
    \begin{tikzpicture}
        \begin{axis}[
            small,
            scale=0.7,
            xtick={1,1.5,2,...,18},
            ytick={0.0,0.01,...,0.9},
            xmin=13, xmax=16,
            ymin=0.71, ymax=0.74,
            xlabel=$\tavg$,
            x label style={at={(axis description cs:.5,.13)},anchor=north},
            y label style={at={(axis description cs:.25,.5)},anchor=south},
            label style={font=\scriptsize},
            tick label style={font=\tiny},
        ]
            \node[] at (axis cs: 13.5,0.716) {\scriptsize $F = 100$};
            \node[] at (axis cs: 13.5,0.713) {\scriptsize $M = 16$};
            \addplot[red] table [x=tavg, y=eff1, col sep=comma] {data/all_100_16_16_20_20_80.csv};
        \end{axis}
    \end{tikzpicture}
    \vskip -.5em
    \end{subfigure}
    \vskip -.7em
    \caption{The zigzag shape of the time efficiency.}
    \label{fig:all_eff2}
    \vskip -.5em
\end{figure}
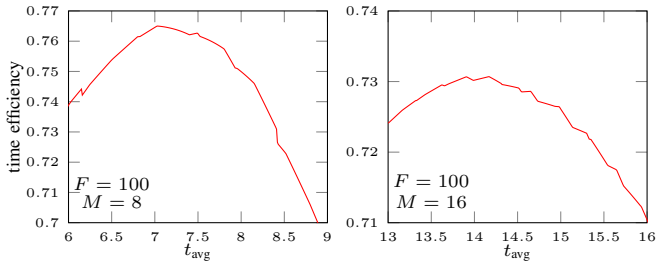

Although the curves for \eqref{eq:eff1}, i.e., the colored curves, look smooth, they are in a zigzag shape.
We zoom the red curves, i.e., $F = 100$, in \Cref{fig:all_eff2} to illustrate this behavior, where the step size for evaluating $\tavg$ is $0.01$.
Each significant drop is due to the decrement of $B$.
We can see that for each $B$, the time efficiency looks linear, which justify the heuristic in \Cref{alg:solve}.

On the other hand, we can see that the optimal time efficiency does not necessarily occur at $\tavg = M$.
\Cref{tab:all_eff} listed the optimal time efficiency and the corresponding $\tavg$ shown in \Cref{fig:all_eff}.
In all of the cases, the optimal $\tavg < M$, which is a counter-intuitive result.

\begin{table}[]
\vskip 1em
    \centering
    \caption{Optimal Time Efficiency}
    \label{tab:all_eff}
    \vskip -0.7em
    \scriptsize
    \begin{tabular}{cc|cccc|c}
         $F$ & $M$ & $B$ & $D/B$ & $\tavg$ & time eff. & upper bound \\ \midrule
         \multirow{2}{*}{100} & 8 & 16 & 1.4697 & 7.03 & 0.7650 & 0.8296\\
         & 16 & 8 & 3.9283 & 14.17 & 0.7307 & 0.8370\\ \midrule
         \multirow{2}{*}{256} & 8 & 39 & 0.7641 & 7.49 & 0.7973 & 0.8296\\
         & 16 & 20 & 2.1837 & 14.65 & 0.7896 & 0.8370\\ \midrule
         \multirow{2}{*}{512} & 8 & 78 & 0.5197 & 7.62 & 0.8105 & 0.8296\\
         & 16 & 39 & 1.4637 & 14.98 & 0.8104 & 0.8370
    \end{tabular}
\end{table}

\section{Concluding Remarks}

We formulated the time efficiency for BATS codes on wireless relay network to maximize the average decoding rate per unit time, i.e., minimizing the decoding time.
Also, we brought out the message that 
it is not optimal to set an average number of recoded packets per batch at the relay node equals the number of packets per batch sent by the source node.

\bibliographystyle{IEEEtran}
\bibliography{ref}

\balance

\end{document}